\documentstyle[12pt]{article}
\begin{document}

\centerline{\bf Future and Origin of our Universe: Modern View}

\medskip

\centerline{A.A. Starobinsky}

\medskip

\centerline{Landau Institute for Theoretical Physics,}
\centerline{Russian Academy of Sciences, Moscow, Russia}

\centerline{and}

\centerline{Isaak Newton Institute for Mathematical Sciences,}
\centerline{University of Cambridge, Cambridge, UK}

\vspace{0.5cm}

\centerline{\bf Abstract}

\medskip
\noindent
The existence of a positive and possibly varying Lambda-term
opens a much wider field of possibilities for the future of our Universe
than it was usually thought before. Definite predictions may be made
for finite (though very large) intervals of time only, as well as in
other branches of science. In particular, our Universe will continue
to expand as far as the Lambda-term remains positive and does not
decay to other forms of matter, even if the Universe is closed. Two
new effects due to the presence of a constant Lambda-term are
discussed: reversal of a sign of the redshift change with time for
sufficiently close objects and inaccessibility of sufficiently distant
objects in the Universe for us. A number of more distant and
speculative possibilities for the future evolution of the Universe is
listed including hitting a space-time singularity during an expansion
phase. Finally, in fantastically remote future, a part of our
Universe surrounding us can become supercurved and superdense due to
various quantum-gravitational effects.

This returns us to the past, to the origin of our Universe from a
superdense state about 14 Gy ago. According to the inflationary
scenario, this state was almost maximally symmetric (de Sitter-like).
Though this scenario seems to be sufficient for the explanation of
observable properties of the present Universe, and its predictions
have been confirmed by observations, the question of the origin of
the initial de Sitter (inflationary) state itself remains open. A
number of conjectures regarding the very origin of our Universe,
ranging from "creation from nothing" to "creation from anything", are
discussed.

\section{Future of the Universe}

It is very popular in cosmology to make definite predictions about
infinitely remote future of our Universe. Such predictions may be
found in virtually any book on cosmology, popular or sophisticated.
Usually they have the following form: \\
1) if the spatial curvature of our Universe is zero or negative,
it will {\it expand eternally}; \\
2) if the spatial curvature is positive, the Universe will stop
expanding in future and begin to {\it recollapse}.

However, it is obvious that any prediction about dynamical evolution
of a physical system cannot remain reliable at infinite time. In any
branch of science, sure forecasts exist for finite periods of time
only, ranging from days in meteorology to millions of years in the
Solar system astronomy. So, how can cosmology be an exception from
this general rule? Evidently, it can't. Therefore, the conviction
that the infinite time prediction given above is reliable should be no
more than an illusion. At present we begin to understand profound
reasons for this.

The impossibility to make exact predictions for infinite time
evolution in cosmology results from the two reasons: 1) absence of
precise knowledge of the present composition of matter in the Universe
and future transformations between different kinds of matter; and
2) imprecise knowledge of present initial conditions for spatial
inhomogeneities in the Universe.
The first reason is vital even for an exactly homogeneous and
isotropic Universe, while the second one requires consideration of
deviations from isotropy and homogeneity. It was thought for a long
time that the second reason is the main source of unpredictability
in remote future, but it seems now that the first reason is the most
important one.

Recent observational data on supernova explosions at high redshifts
$z\sim 1$~ obtained by two groups independently~\cite{Perl,Garn}, as
well as numerous previous arguments (see, e.g.,~\cite{Kof85,St94}),
strongly support the existence of a new kind of matter in the Universe
which energy density is positive and dominates over energy densities
of all previously known forms of matter. This form of matter has a
strongly negative pressure and remains unclustered at all scales where
gravitational clustering of baryons and cold non-baryonic dark matter
is seen. Its gravity results in an acceleration of the expansion of
the present Universe: $\ddot a(t_0) >0$, where $a(t)$ is the scale
factor of the Friedmann-Robertson-Walker (FRW) isotropic cosmological
model with time $t$ measured from the cosmological singularity
(the Big Bang) in the past, $t_0$ is the present moment. In the first
approximation, this kind of matter may be described by a constant 
Lambda-term in gravity equations which was introduced by
Einstein. However, a Lambda-term (also called quintessence sometimes)
might be slowly varying with time. If so, this will be soon determined
from observational data. In particular, if we use the simplest model of
a variable Lambda-term borrowed from the inflationary scenario of the
early Universe, namely, an effective scalar field $\phi$ with some
self-interaction potential $V(\phi)$ minimally coupled to gravity,
then the functional form of $V(\phi)$ may be determined from
observational {\em cosmological functions}: either from the luminosity
distance $D_L(z)$~\cite{St98,HT}, or from the linear density
perturbation in the dust-like (cold dark matter (CDM) plus baryon)
component of matter in the Universe ${\delta \rho \over \rho}(z)$~
\cite{St98} (provided the Lambda-term satisfies
the weak energy condition $\varepsilon_{\Lambda}+p_{\Lambda}\ge 0$).

Should the Lambda-term be always exactly constant, the prediction
for the future of the Universe is simple and boring: the
Universe will expand forever, energy densities of all kinds
of matter apart from the Lambda-term tend to zero exponentially, and
the space-time metric locally approaches the de Sitter metric
(though globally it has a much more general quasi-de Sitter form,
see~\cite{St83}). Thus, in this case the Universe becomes cold and
empty finally. However, this is just the point: we are {\em not
sure} that the Lambda-term will remain exactly the same at all times.
And if it changes with time, predictions for remote future of the
Universe may appear completely different.

On the other hand, sure forecasts for finite intervals of time
are certainly possible in cosmology. Moreover, it is the present
high degree of order in the Universe that makes the interval
of predictability very large - much larger than in other branches
of science. By the way, let us note that according to the
inflationary scenario the present regularity of the Universe
is a consequence of the fact that the Universe was even more
regular - actually, almost maximally symmetric - in the past,
during a de Sitter (inflationary) stage. The curvature at that
stage was very high, close to the Planck curvature (though at
least five orders of magnitude less near the end of the
inflationary stage), in sharp contrast with a very low curvature
at the asymptotic quasi-de Sitter stage in future discussed in the
previous paragraph. Let me give you an example of such kind of
predictions. If we make the following three assumptions: the present
Hubble constant $H_0\ge 50$ km s$^{-1}$ Mpc$^{-1}$, the present
age of the Universe $t_0\ge 10$ Gy and the energy density of the
Lambda-term is non-negative (and will remain so for the period of
time given below), than the Universe will continue its expansion
for at least 20 Gy irrespective of the sign of its spatial curvature~
\cite{St88}. At present, we are practically sure from existing
observational data that all these three assumptions are correct.
Since this interval exceeds the time of active life of main sequence
stars (and the Sun, in particular), this estimate is more than
sufficient for discussion of the future of the Earth and
human civilization.

Derivation of this result goes as follows. If $\varepsilon_{\Lambda}
\ge 0$, the most critical case with respect to recollapse of the 
Universe in future occurs just when $\varepsilon_{\Lambda}\equiv 0$
and the Universe is closed (${\cal K}=1$, positive spatial curvature).
The law of the evolution of a closed dust-dominated FRW cosmological
model has the following parametric form:
\begin{equation}
a={1\over 2}a_{max}(1-\cos\eta)~,~~t={1\over 2}a_{max}(\eta - \sin
\eta)~,~~0\le \eta \le 2\pi~,
\end{equation}
where $a_{max}$ is the maximal radius of the Universe (I put $c=1$ 
here and below). The parameter $\eta$ is the conformal time
$\eta =\int dt/a(t)$ actually.

The corresponding Hubble parameter is
\begin{equation}
H(t)\equiv {d\over dt}\ln a(t)= {2\over a_{max}}{\sin \eta \over 
(1-\cos\eta)^2}~.
\end{equation}
Note that the Hubble constant $H_0=H(t_0)$. Then it follows from the
inequalities for $H_0$ and $t_0$ given above that
\begin{equation}
H_0t_0={\sin \eta_0~(\eta_0-\sin\eta_0)\over (1-\cos\eta_0)^2}
\ge 0.51~, ~~\eta_0\le 1.92
\label{etaineq}
\end{equation}
where $\eta_0=\eta(t_0)$. The remaining time of expansion before 
beginning of recollapse of the Universe which takes place
at $\eta = \pi$ in this model is:
\begin{equation}
T_{exp}={\pi \over 2}a_{max}-t_0=t_0~{\pi -\eta_0+\sin\eta_0\over
\eta_0 - \sin\eta_0}\ge 2.2t_0\ge 22~{\rm Gy}~.
\label{time}
\end{equation}
Given above was just the rounded form of this inequality. 
Incidentally, it follows from (\ref{etaineq}) that the upper limit
on the present energy density of dust-like matter in terms of the 
critical one $\varepsilon_c=3H_0^2/8\pi G$ is
$\Omega_m=\varepsilon_m/\varepsilon_c \le 1.5$. Of course, 
presently existing observational data, especially the supernova data 
mentioned above and data on temperature angular anisotropy 
${\Delta T\over T}$ of the cosmic microwave background (CMB) restrict
spatial curvature of the Universe even better: $|\Omega_m+
\Omega_{\Lambda}-1|\le 0.3$ (see, e.g., the second reference in~
\cite{Garn}, and~\cite{Teg}). 

Still people are interested in more and more remote future.
Predictions for this period can be made, of course, but they become
less and less reliable with time growth, because we have to base on
more and more assumptions. So, speaking about very remote future, we
can at best present a list of some possibilities for future evolution
of the Universe. This list, however incomplete it is, shows that real
future evolution of the Universe is infinitely complicated and has
no boring smooth asymptotic behaviour at $t\to \infty$.

But before discussing these remote possibilities, let me mention two
significantly new effects which arise in the case of a constant
$\Lambda$-term ($\varepsilon_{\Lambda} >0$). From now on, I assume
that the Universe is spatially flat (${\cal K}=0$) for the following
reasons:~a) no observational data directly point to ${\cal K}\not= 0$
at present;~b) a spatial curvature of the Universe is strongly bounded
as mentioned above, and does not dominate over matter (including both
dust-like matter and a $\Lambda$-term);~c) the simplest inflationary
models of the early Universe predict $|\Omega_m+\Omega_{\Lambda}-1|
\ll 1$;~d) for simplicity.

\medskip

1. Reversal of a sign of $\dot z$ for sufficiently close objects.

\medskip 

Let us consider the question how the redshift of a given object changes 
with time. The present redshift $z\equiv z(t_0)$ is given by the 
expression 
\begin{equation}
1+z={a(\eta_0)\over a(\eta_{em})},~~\eta_{em}=\eta_0-r~,
\label{z}
\end{equation}
where $r$ is the constant coordinate (comoving) distance to the object
and $\eta_{em}=\eta (t_{em})$ is the moment when the object emitted light 
observing now. The physical distance to the object is $R=ar$. To find
$\dot z$, one has to differentiate (\ref{z}) with respect to $t_0$.
If $\Lambda=0$, then $\dot z<0$ for all $z$. Moreover, $z(t)$
monotonically decreases with time and tends to $0$ as $t\to \infty$.
On the contrary, if $\Lambda > 0$, $z(t)$ stops decreasing at some moment
and then begin to increase due to an acceleration of the Universe in the
$\Lambda$-dominated regime. As a result, $\dot z>0$ if $z<z_c$ at the 
present time. The value $z_c$ for which $\dot z_c(t_0)=0$ (so 
$\dot z$ considered as a function of $z$ for given the $t=t_0$ changes its 
sign) is determined from the equation:
\begin{equation}
\dot a(t_0)=\dot a(t_{em}(z_c))~, ~~\eta_{em}(z_c)=\eta_0-r(z_c)~.
\end{equation}
If the Universe is flat, then this equation reduces to the algebraic
equation
\begin{equation}
(1+z_c)\left(\Omega_m+{1-\Omega_m\over (1+z_c)^3}\right)=1~.
\end{equation}
In particular, $z_c=2.09$ if $\Omega_m=0.3$ which is the best fit to the 
supernova data~\cite{Perl, Garn}. Note that $z_c$ decreases with
increasing $\Omega_m$. This effect may be even directly observed in
future, though not too soon because measuring $\dot z$ represents a 
formidable task (see the discussion of problems arising in~\cite{Loeb}).

\medskip

2. Loss of possibility to reach distant objects. 

\medskip

The existence of a constant $\Lambda >0$ leads to the appearance of the 
future event horizon (as in the de Sitter space-time). This means that
looking at sufficiently remote galaxies with $z>z_{eh}$ at the present 
time, we can neither reach them physically in an arbitrary long time 
period,
nor even send a message to intelligent beings in them (supposing that
such exist or will appear in future) saying ``we are!''. In other words,
the coordinate volume of space which our civilization may affect is
finite. Its border is given by $r_{eh}=\eta(t=\infty)-\eta_0.$ The
redshift $z_{eh}(r_{eh},\Omega_m)$ can found from the equation
\begin{equation}
\int_1^{1+z_{eh}}{dx\over \sqrt{1-\Omega_m+\Omega_mx^3}}=
\int_0^1 {dx\over \sqrt{1-\Omega_m+\Omega_mx^3}}
\end{equation}
(both sides of this equation are equal to $R_{eh}H_0=a(t_0)r_{eh}H_0$).
If $\Omega_m=0.3$, then $z_{eh}=1.80$ (note that $z_{eh}$ grows with 
$\Omega_m$ reaching infinity for $\Omega_m=1$). This is not much, we see 
many galaxies and quasars with larger redshifts. So, all of them are
unaccessible for us. Another similar effect was recently considered 
in~\cite{STV}.

\medskip

Now we return to long-time predictions. The standard one usually 
presented refers to the case of a constant $\Lambda >0$. Then, as was 
already mentioned above, the Universe will expand infinitely for
any sign of its spatial curvature. It quickly approaches the de
Sitter state with $H=H_{\infty}=\sqrt{\Lambda/3}=H_0\sqrt{1-\Omega_m}$. 
So, this scenario may be called ``inflation in future''.
Matter density $\varepsilon_m \propto a^{-3}(t)\to 0$ while
density perturbations $\delta\varepsilon_m/\varepsilon_m \to const$
if they are still in the linear regime now. Circumstantially, CMB
multipole angular anisotropies $(\Delta T/T)_l$, in particular the
quadrupole one, freeze at some constant values, too (see the first
reference in~\cite{Kof85}). On the other hand, gravitationally bound 
systems which physical size is $R<10h^{-1}$ Mpc at present (our Galaxy, 
in particular) will remain bound, at least as far as classical gravity
is concerned (here $h=H_0/100$ km s$^{-1}$ Mpc$^{-1}$). So, islands 
of galaxies will remain in the ever expanding and becoming more and 
more vacuum-like on average Universe.

However, this is not the only possibility for a future fate of the
Universe even at the classical level, and probably not the correct one 
at all if quantum-gravitational effects are taken into account.
A number of possible alternatives is presented below.

\medskip

1. Decay of $\Lambda$ in future.

\medskip

If a $\Lambda$-term is unstable and decays faster than $a^{-2}$
(i.e., $\varepsilon_{\Lambda}a^2\to 0$ at $t\to \infty$), then
recollapse of some parts of the Universe becomes possible due to
existing inhomogeneities even if ${\cal K}=0$. A $\Lambda$-term
may decay with time, e.g., in the simplest scalar field model 
mentioned above if $V(\phi)$ decreases sufficiently fast with 
growth of $a(t)$. At present, the $\Lambda$-term is changing 
rather slowly, if at all. If we assume for simplicity
that its pressure $p_{\Lambda}=k\varepsilon_{\Lambda},~k=const$,
then it follows from observational data that $k<-0.6$ (see, e.g.,
\cite{Efst}). Since $\varepsilon_{\Lambda}\propto a^{-3(1+k)}$
in this case, this corresponds to $\varepsilon_{\Lambda}$ decaying
less rapidly than $a^{-1.2}$ at present. However, this behaviour may 
change in future.

\medskip

2. Collision with a null singularity.

\medskip

There exists a rather unpleasant possibility that our future world
line will cross a real space-time singularity with infinite
values of the Riemann tensor (though its scalar invariants are less
singular and may even remain finite sometimes) concentrated at a null
hypersurface. So, this singularity may be called a gravitational shock
wave with an infinite amplitude. It was conjectured that such 
singularities should arise
along Cauchy horizons inside rotating or charged black holes~\cite{Isr},
and it has been shown that this really occurs in some simplified cases
(see~\cite{Burko} for the most recent treatment).

It not is clear at present if this collision is deadly to an intelligent
life. However, it is certainly fatal for our ability to predict future
of our Universe since any classical extension of space-time beyond
such a singularity is non-unique. The most unpleasant is the fact
that an intelligent being cannot even forecast this event until
the shock wave hits him/her. Fortunately, this possibility seems to be
rather improbable since it requires a very specific global space-time 
structure of the Universe (namely, the existence of a Cauchy horizon
intersecting our future light cone). However, I cannot exclude it
completely basing on our present knowledge.

\medskip

3. Formation of a classical space-like curvature singularity
during expansion.

\medskip

To hit a real space-time singularity with infinite invariants of the 
Riemann tensor, it is not necessary to have an isotropic recollapse 
first. Such a singularity may also occur as a result of sudden growth
of anisotropy and inhomogeneity at some moment during expansion, or 
even as a result of infinite growth of $a(t)$ in a finite time period.
The former possibility realizes, e.g., in the model of a variable 
$\Lambda$-term based on a scalar field with a self-interaction
potential $V(\phi)$ as before, but non-minimally coupled to gravity  
due to the term $\xi R\phi^2$ in its Lagrangian density. If $\xi >0$ and 
if the field $\phi$ will reach the critical value $\phi_{cr}=
1/\sqrt{8\pi\xi G}$ at some finite moment of time $t_{cr}$ in future, the
effective gravitational constant $G_{eff}$ becomes infinite, small 
spatial inhomogeneities grow without limit and a generic inhomogeneous
space-like singularity (not oscillating) forms~\cite{St81}. Very close
to this singularity, the volume factor $\sqrt{-g}$ stops growing
and finally approaches zero $\propto (t_{cr}-t)^q,~0<q<1$, but this 
recollapse is strongly anisotropic.

The latter possibility takes place in an even simpler case (though
not justified by a reasonable field-theoretic model) of the linear 
equation of state $p_{\Lambda}=k\varepsilon_{\Lambda},~k=const$
with $k<-1$, so that the weak energy condition $p_{\Lambda}+
\varepsilon_{\Lambda}\ge 0$ is violated at the classical level.
Then $a(t)$ becomes infinite (and the curvature singularity is reached) 
in a finite interval of time (measured from the present moment)
\begin{equation}
T_s=H_0^{-1}{2\over 3|1+k|}\int_0^1{dx\over \sqrt{1-\Omega_m
+\Omega_m x^{{2|k|\over |1+k|}}}}~.
\end{equation}
As was discussed above, the $\Lambda$-term is changing sufficiently
slowly, if at all. Using the supernova data, it can be shown that $k$ 
should be certainly more than $-1.5$. Then, taking $\Omega_m=0.3$ and
$H_0=70$ km s$^{-1}$ Mpc$^{-1}$, we obtain $T_s>22$ Gy. So, even for 
this very speculative model, we get practically the same lower bound 
on the period of safe expansion of Universe in future as was given 
before in Eq. (\ref{time}).

More justified and refined field-theoretic models having such a regime
which is called ``superinflation'', or ``pole inflation'' do exist.
In particular, this regime was already present among possible solutions
of the higher-derivative gravity model used in~\cite{St80} to
construct the first viable cosmological model of the early Universe
with the initial de Sitter (inflationary) stage (though, of course, 
another solution of this model having the ``graceful exit'' from inflation
to the FRW radiation-dominated stage was used in this paper). Another model 
where pole inflation occurs is the ``Pre-Big-Bang'' scenario of the early
Universe~\cite{Ven}. So, could a ``Post-Big-Bang'' in future be possible? 
Once more, I cannot exclude this possibility now. 

\medskip

4. Hitting a space-like singularity in future due to 
quantum-gravitational effects.

\medskip

Finally, if none of the classical effects listed above (and other ones
not known now) occurs, there always exist quantum-gravitational 
fluctuations. They are non-trivial (not coinciding with vacuum fluctuations
in the Minkowski space-time) if $\Lambda\not= 0$. There are two kinds of 
them.

\medskip

A. Fluctuations of an effective scalar field producing a $\Lambda$-term.

\medskip

During future expansion of the Universe at the $\Lambda$-dominated stage,
these fluctuations may occasionally result in jumps to a higher energy
(and a higher curvature) state (``false vacuum''), in particular, even to 
an initial inflationary state. Depending on an effective mass of this
scalar field, this transition may occur either in one jump~\cite{Wein},
see also recent papers~\cite{GV} (where this process was called ``recycling
of the Universe'') and~\cite{Rub}, or as a result of a long series of small
jumps, as it occurred during stochastic inflation in the early Universe~
\cite{St82,St86}. So, in the latter case we have ``stochastic inflation in 
future''. 
 
In both cases, it is necessary that the whole part of the Universe
inside the de Sitter event horizon (or even a little bit larger) makes 
this transition. It is clear that the probability of this process is 
fantastically small. I don't think that one can really grasp how small 
it is by his/her senses. Still it is non-zero, so this event will occur 
finally. This probability mainly depends on the future asymptotic value 
of a $\Lambda$-term $\Lambda_{\infty}=3H_{\infty}^2$:
\begin{equation}
w_s\sim \exp \left( {\pi\over GH_f^2}-{\pi\over GH_{\infty}^2}\right)~,
\label{ws}
\end{equation}
where $H_f^2=\Lambda_f/3$ is the curvature of a false vacuum state. The
second term in the exponent is $\sim 10^{122}$, so it is practically
impossible to imagine how large is a typical time required for this 
transition. However, it is finite. Thus, in this case future curvature 
space-like singularity is reached during continuous expansion of the 
Universe.

\medskip

B. Quantum fluctuations of the gravitational field.

\medskip

However, it appears that it is much simpler to reach future curvature
singularity due to quantum fluctuations of the gravitational field itself.
These fluctuations can produce a significant anisotropy
described by a non-zero value of the conformal Weyl tensor comparable
to that of the Riemann tensor. The corresponding quantum transition
may be described by the $S_2\times S_2$ instanton:
\begin{eqnarray}
ds^2 = d\tau^2+H_1^{-2}\sin^2H_1\tau~dx^2+ H_1^{-2}d\Omega^2 
= (1-H_1^2\tilde x^2)~d\tilde\tau^2 + {d\tilde x^2 \over 1-H_1^2
\tilde x^2}+ H_1^{-2}~d\Omega^2~, \\
d\Omega^2=d\theta^2+\sin^2\theta~d\varphi^2~,~~H_1^2=\Lambda_{\infty}=
3H_{\infty}^2~. \nonumber
\end{eqnarray}
Here $\tilde \tau$ is a cyclic variable with the period $2\pi/H_1$. The 
second, ``thermal'' form of the instanton suggests that the transition 
occurs in a ``local'' part of the Universe with a size slightly larger than 
$H_1^{-1}$. The resulting space-time metric after the transition is:
\begin{equation}
ds^2= (1-H_1^2\tilde x^2)~d\tilde t^2 - {d\tilde x^2 \over 1-H_1^2
\tilde x^2}- H_1^{-2}~d\Omega^2~,
\end{equation}
which covers a part of the Bondi-Nariai space-time~\cite{BN} with a finite 
range of $x$:
\begin{equation}
ds^2=dt^2-a^2(t)~dx^2-b^2(t)~d\Omega^2~,~~ a(t)=H_1^{-1}\cosh H_1t~,
~~b=H_1^{-1}=const~, 
\label{BNmet}
\end{equation}
(see~\cite{KSS} for discussion of quantum-gravitational 
effects in the metric (\ref{BNmet})). Note that the choice $a(t)=
a_1\exp H_1t$ is also possible. It corresponds to covering of
another part of the Bondi-Nariai space-time.

The probability of this quantum jump is given by the difference of actions
for the $S_4$ and $S_2\times S_2$ instantons with the same value of 
$\Lambda$:
\begin{equation}
w_g\sim \exp \left(-{\pi\over GH_1^2}\right)~.
\label{wg}
\end{equation}
Note that the exponent in Eq. (\ref{wg}) is $3$ times less by modulus
than that in Eq. (\ref{ws}). Thus, this second process due to purely
quantum-gravitational fluctuations is much more probable, $w_s \sim
w_g^3$ (though, of course, $w_g$ is fantastically small, too).   

What happens with the considered region of space-time after the jump?
The space-time (\ref{BNmet}) is classically unstable with respect to
long-wave gravitational perturbations ($\Lambda =const$). With the 
probability $0.5$, $b$ grows up and then this region returns to
the locally de Sitter behaviour $a(t)\propto b(t)\propto \exp 
(H_{\infty}t)$ at $t\to \infty$ (so that the whole space-time
approaches a specific form of the general quasi-de Sitter 
asymptote~\cite{St83}). On the other hand, with the other $0.5$ 
probability, $b$ goes down, the region begins to recollapse soon, and 
the Kasner singularity $a(t)\propto (t_1-t)^{-1/3},~b(t)\propto 
(t_1-t)^{2/3}$ forms. Thus, this region of the Universe returns to a 
supercurved state.

\medskip

So, one way or another, local parts of the Universe return to a singular
supercurved state, though it might require a very huge amount of time.
Thus, it seems at present that ``cold death'' is not a viable possibility
for the future of our Universe. Let me emphasize that this return to
a future singularity occurs in a very inhomogeneous fashion in all
examples considered above. Therefore, any finite coordinate volume
of the Universe becomes more and more inhomogeneous with time growth,
in accordance with the Second Law of thermodynamics (understood in
a very broad and imprecise sense). The same refers to the global 
structure of the Universe: it becomes more and more complicated in future,
too. On the other hand, characteristic times for significant growth
of complexity of our Universe are very large. As a result, the Universe will
certainly remain very ordered for periods of the order of a few tenths
of Gy that significantly exceeds its present age.

What happens after the return to a singular state? We don't know it
at the present state of the art. Still it is possible to conjecture
that at least a very small part of the region which hits a singularity
will bounce back and return to a low-curvature state. Especially 
interesting and remarkable would be if, during the process, this part
spend some time at an inflationary stage. Then infinitely many low 
curvature and ordered universes similar to our present Universe may be 
created from this part in future. Repeating all this hypothetical, but
not firmly prohibited process more and more, we 
see that the future of our Universe may be not simply {\em very}
complicated but even {\em infinitely} complicated.

\section{Past of the Universe}

We see that discussion of the future of our Universe has naturally
led us to the question of the origin of our Universe in the past, about
$14$ Gy ago. The preferred and very well developed theory of a period
of the evolution of the Universe preceding the hot radiation-dominated
FRW stage is given by the inflationary scenario of the early Universe.
According to this scenario, our Universe was in an almost maximally 
symmetric (de Sitter, or inflationary) state during some period of time
in the past. I think that the main attractive features of the inflationary 
scenario are the following: 1) its extreme aesthetic elegance and beauty,
and 2) complete predictability of properties of the observed part of the
Universe after the end of the inflationary stage (in particular, at the 
present time). Thus, any concrete realization of the inflationary
scenario may be falsified by observations, and many of them had been
falsified already.

But it is remarkable that there exist a large class of the so called
simplest inflationary models (with one slowly rolling effective scalar 
field producing the inflationary stage) whose predictions, just the 
opposite, were confirmed by observations. This especially refers to 
results of a COBE satellite experiment where low multipoles of the CMB
angular temperature anisotropy $(\Delta T/T)_l$ with $l$ up to $\sim 20$
were measured, and to results of numerous recent medium- and small-angle
measurements of $\Delta T/T$ which confirm the inflationary prediction 
about the location and the approximate height of the so called first
acoustic (Doppler) peak. So, the inflationary scenario really has a
large predictive power!

Still it is clear that since any inflationary stage is not stable, but 
only metastable, it cannot be the {\em very} beginning of our Universe.
Something was before, that was the origin of the inflationary stage.
The most well known proposal, put forward long before the inflationary 
scenario was introduced in 1979-1982, was the ``creation of the Universe
from nothing''~\cite{TF}. Here nothing means literally nothing, in 
particular, that were no space-time before our Universe was created.
This idea does not work without some inflationary state following
the creation, so it was forgotten for some time and was revived~\cite{GZ}
only after the development of the inflationary scenario. In that case the
creation is mathematically described by the $S_4$ (de Sitter) instanton.
In the papers~\cite{TF,GZ} the creation of a closed FRW universe was
considered, however, it was recently shown that an open FRW universe
may be produced ``from nothing'', too, using approximately the same
(though already singular) instanton~\cite{HawTur}.

However, at the same moment the idea of ``creation from nothing'' was
renewed, it was pointed that this is not the {\em only} possibility
to create an inflationary stage~\cite{St82a}. Let me present an incomplete 
list of other alternatives. \\
1.~Quasi-classical motion of space-time from a generic inhomogeneous
anisotropic singularity to the de Sitter attractor solution. \\
2.~Decay of less symmetric, higher curvature self-consistent solutions
of gravity equations with all quantum corrections included (e.g., the 
Bondi-Nariai solution (\ref{BNmet})). \\
3.~Stochastic drift from a singularity with the Planckian value of 
curvature
along a sequence of de Sitter-like solutions (this is what actually
occurs in the so called eternal chaotic inflation~\cite{Lin}).\\
4.~Quantum nucleation of our Universe from some other ``Super-universe'',
in particular, even from some asymptotically flat space-time (the latter
possibility includes ``creation of the Universe in a laboratory'',
see~\cite{FG}). \\
5.~Creation of the Universe from a higher-dimensional space-time. \\
Evidently, many more possibilities remain not mentioned. It seems that
they are all indistinguishable from observations. That is why, in order
to tackle this great ambiguity, a completely different principle
of ``creation of the Universe from anything'' was put forward 
in~\cite{SZ88}. Namely, it states that: \\
{\it ``local'' observations cannot help distinguish between different 
ways of formation of an inflationary stage}. \\
By ``local'' I mean all observations inside the presently observed
Universe, and even all observations made along our future world line
in arbitrary remote future. ``Creation from anything'' intrinsically
includes all ways of creating the de Sitter (inflationary) stage,
with the ``creation from nothing'' being only one (and therefore,
scarcely probable) way among them.

It is amusing that the mathematical description of ``creation from 
anything'' is based on the same $S_4$ instanton as ``creation from 
nothing'', but now written in a static, ``thermal'' form:
\begin{equation}
ds^2=(1-H^2r^2)~d\tau^2 + {dr^2\over 1-H^2r^2} + r^2d\Omega^2~,
\end{equation}
where $\tau$ is periodic with the period $2\pi/H$ -- the inverse 
Gibbons-Hawking temperature~\cite{GH} (I assume here 
$\Lambda=const =3H^2$ for simplicity).

Now, using the thermal interpretation of the $S_4$ instanton, we may
ascribe the total entropy 
\begin{equation}
S({\rm entropy})=|S|({\rm action})= {\pi \over GH^2}\gg 1
\label{S}
\end{equation}
to the Universe at the inflationary stage. This entropy just reflects the 
absence of knowledge of a given observer about a space-time structure beyond
the de Sitter horizon and about a way how this de Sitter stage was
formed. Since $\sqrt{G} H<10^{-5}$ at the end of an inflationary stage,
$S>10^{10}$ there.

Of course, this principle (as all principles introduced by hand) may be a 
little bit extreme. I cannot exclude the possibility that we shall be able
to get some knowledge about a pre-inflationary history of our Universe. Then
a value of the entropy of the Universe at the end of an inflationary stage
will be less than that given by Eq. (\ref{S}).

\vfill

\begin{thebibliography}{99}
\bibitem{Perl} S.~Perlmutter, G.~Aldering, M.~Della Valle
{\it et al.}, Nature {\bf 391}, 51 (1998); S.~Perlmutter, G.~Aldering,
G.~Goldhaber {\it et al.}, Astroph. J. {\bf 517}, 565 (1999).
\bibitem{Garn} P.M.~Garnavich, R.P.~Kirshner, P.~Challis {\it et al.},
Astrophys. J. Lett. {\bf 493}, L53 (1998); A.G.~Riess, A.V.~Philipenko,
P.~Challis {\it et al.}, Astron. J. {\bf 116}, 1009 (1998).
\bibitem{Kof85} L.A.~Kofman and A.A.~Starobinsky, Sov. Astron. Lett.
{\bf 11}, 271 (1985); L.A.~Kofman, N.Yu.~Gnedin, and N.A.~Bahcall,
Astroph. J. {\bf 413}, 1 (1993); J.P.~Ostriker and P.J.~Steinhardt,
Nature {\bf 377}, 600 (1995); J.S.~Bagla, T.~Padmanabhan, and
J.V.~Narlikar, Comm. Astrophys. {\bf 18}, 275 (1996).
\bibitem{St94} A.A.~Starobinsky, in {\it Cosmoparticle Physics. I.
Proceedings of the 1st International Conference on Cosmoparticle
Physics "Cosmion-94"}, Moscow, 5-14 Dec. 1994, eds. M.Yu.~Khlopov,
M.E.~Prokhorov, A.A.~Starobinsky, and J.~Tran Thanh Van, Edition
Frontiers, 1996, p. 141 (e-mail preprint archive astro-ph/9603074).
\bibitem{St98} A.A.~Starobinsky, JETP Lett. {\bf 68}, 757 (1998).
\bibitem{HT} D.~Huterer and M.S.~Turner, Phys. Rev. D, in press (1999)
(e-mail preprint archive astro-ph/9808133); T.~Nakamura and T.~Chiba, 
Mon. Not. Roy. Ast. Soc. {\bf 306}, 696 (1999).
\bibitem{St83} A.A.~Starobinsky, JETP Lett. {\bf 37}, 66 (1983).
\bibitem{St88} A.A.~Starobinsky. {\it The Universe}, in {\it
Physical Encyclopedia}, Vol. I, Moscow, Soviet Encyclopedia, 1988,
p. 346 (in Russian).
\bibitem{Teg} M.~Tegmark, Astroph. J. Lett. {\bf 514}, L69 (1999).
\bibitem{Loeb} A.~Loeb, e-mail preprint archive astro-ph/9802122 (1998).
\bibitem{STV} G.~Starkman, M.~Trodden, and T.~Vachaspati, Phys. Rev. 
Lett. {\bf 83}, 1510 (1999).
\bibitem{Efst} G.~Efstathiou, e-mail preprint archive astro-ph/9904356
(1999).
\bibitem{Isr} E.~Poisson and W.~Israel, Phys. Rev. D {\bf 41}, 1796
(1990); A.~Ori, Phys. Rev. Lett. {\bf 67}, 789 (1991); {\it ibid}~
{\bf 68}, 2117 (1992). 
\bibitem{Burko} A.~Ori, Phys. Rev. {\bf D57}, 4745 (1998); L.~M.~Burko,
Phys. Rev. {\bf D60}, 104033 (1999).
\bibitem{St81} A.A.~Starobinsky, Sov. Astron. Lett. {\bf 7}, 36 (1981).
\bibitem{St80} A.A.~Starobinsky, Phys. Lett. {\bf 91B}, 99 (1980).
\bibitem{Ven} G.~Veneziano, Phys. Lett. {\bf 265B}, 287 (1991); 
M.~Gasperini and G.~Veneziano, Astropart. Phys. {\bf 1}, 317 (1993).
\bibitem{Wein} K.~Lee and E.J.~Weinberg, Phys. Rev. {\bf D36}, 1088 (1987).
\bibitem{GV} J.~Garriga and A.~Vilenkin, Phys. Rev. {\bf D57}, 2230 (1998).
\bibitem{Rub} V.A.~Rubakov and S.M.~Sibiryakov, e-mail preprint archive
gr-qc/9905093.
\bibitem{St82} A.A.~Starobinsky, Phys. Lett. {\bf 117B}, 175 (1982).
\bibitem{St86} A.A.~Starobinsky, in {\it Field Theory, Quantum Gravity and 
Strings}, ed. H.J.~de Vega and N.~Sanchez, Lect. Notes in Physics
(Springer-Verlag) {\bf 246}, 107 (1986). 
\bibitem{BN} H.~Bondi, Mon. Not. Roy. Astron. Soc. {\bf 107}, 410 (1947);
H.~Nariai. Sci. Rep. Tohoku Univ. {\bf 34}, 160 (1950); {\it ibid}~
{\bf 35}, 62 (1951).
\bibitem{KSS} L.A.~Kofman, V.~Sahni, and A.A.~Starobinsky, JETP {\bf 58},
1090 (1983). 
\bibitem{TF} E.P.~Tryon, Nature {\bf 246}, 396 (1973); P.I.~Fomin, Dokl.
Akad. Nauk Ukr. SSR {\bf A9}, 831 (1975).
\bibitem{GZ} L.P.~Grishchuk and Ya.B.~Zeldovich., in {\it Quantum
Structure of Space-Time}, ed. M.~Duff and C.I.~Isham, Camb. Univ. Press,
1982, p. 409.
\bibitem{HawTur} S.W.~Hawking and N.~Turok, Phys. Lett. {\bf 425B}, 25 (1998).
\bibitem{St82a} A.A.~Starobinsky, in {\it Proc. of the Second Seminar
``Quantum Theory of Gravity'' (Moscow, 13-15 Oct. 1981)}, INR Press,
Moscow, 1982, p. 58; reprinted in {\it Quantum Gravity}, ed. M.A.~Markov
and P.C.~West, Plenum Publ. Co., New York, 1984, p. 103.
\bibitem{Lin} A.D.~Linde, Mod. Phys. Lett. {\bf A1}, 81 (1986); Phys. Lett.
{\bf 175B}, 395 (1986). 
\bibitem{FG} E.~Farhi and A.H.~Guth, Phys. Lett. {\bf 183B}, 149 (1987);
E.~Farhi, A.H.~Guth, and J.~Guven, Nucl. Phys. {\bf B339}, 417 (1990).
\bibitem{SZ88} A.A.~Starobinsky and Ya.B.~Zeldovich, {\it The
Spontaneous Creation of the Universe}, in {\it Sov. Sci. Rev. E -
Astroph. Space Phys.}, ed. R.A.~Syunyaev (Harwood Academic Press, 
New York), {\bf 6}, part 2, 103 (1988).
\bibitem{GH} G.W.~Gibbons and S.W.~Hawking, Phys. Rev. {\bf D15}, 2738
(1977). 
\end{thebibliography}
\end{document}